\documentclass[twocolumn,amsmath,amssymb]{revtex4}

\usepackage{graphicx}
\usepackage{dcolumn}
\usepackage{bm}

\DeclareMathOperator{\erf}{erf}
\DeclareMathOperator{\erfgau}{erfgau}
\newcommand{\Psimu}{\ensuremath{\Psi^{\mu}}}
\newcommand{\bra}[1]{\ensuremath{\langle #1 \vert}}
\newcommand{\ket}[1]{\ensuremath{\vert #1  \rangle}}

\renewcommand{\b}[1]{\ensuremath{\mathbf{#1}}}

\begin{document}

\title{Exchange-correlation potentials and local energies per particle along non-linear adiabatic connections}

\author{Julien Toulouse}
\author{Fran\c{c}ois Colonna}
\author{Andreas Savin}
 \email{savin@lct.jussieu.fr}
\affiliation{
Laboratoire de Chimie Th\'eorique, CNRS et Universit\'e Pierre et Marie Curie,\\
4 place Jussieu, 75252 Paris, France
}

\date{\today}
             
\begin{abstract}
We study non-linear adiabatic connection paths in density-functional theory using modified electron-electron interactions that perform a long-range/short-range separation of the Coulomb interaction. These adiabatic connections allows to define short-range exchange-correlation potentials and short-range local exchange-correlation energies per particle that we have calculated accurately for the He and Be atoms and compared to the corresponding quantities in the local density approximation (LDA). The results confirm that the LDA better describes exchange-correlation potentials and local exchange-correlation energies per particle when the range of the interaction is reduced.
\end{abstract}

\maketitle

\section{Introduction}
\label{sec:Intro}

The adiabatic connection procedure of density functional theory (DFT)~\cite{HohKoh-PR-64} connects the fictitious non-interacting Kohn-Sham (KS)~\cite{KohSha-PR-65} system to the physical one by continuously switching on the electron-electron interaction. The connection is carried out at constant density and is controlled by a interaction parameter $\mu$.  The partially interacting $N$-electron system for a given value of $\mu$ along this connection is described by the Hamiltonian
\begin{equation}
\hat{H}^{\mu} = \hat{T} + \hat{V}_{ee}^{\mu} + \hat{V}^{\mu},
\label{Hmu}
\end{equation}
where $\hat{T}$ is the kinetic energy operator, $\hat{V}_{ee}^{\mu}=\sum_{i<j}v_{ee}^{\mu}(r_{ij})$ is the modified electron-electron interaction operator and $\hat{V}^{\mu}=\sum_{i} v^{\mu}(\b{r}_i)$ is the corresponding local external potential which ensures that this fictitious system has the same ground-state density $n$ than that of the physical system. The ground-state wave function of this partially interacting system is denoted by $\Psimu$.

In the more common adiabatic connection (see, e.g., Ref.~\onlinecite{Har-PRA-84}), the electron-electron interaction is switched on linearly. However, other non-linear paths are possible~\cite{Yan-JCP-98}. In particular, in view of constructing a multi-determinantal extension of the KS scheme of DFT, it has been proposed~\cite{StoSav-INC-85,Sav-INC-96a,Sav-INC-96,LeiStoWerSav-CPL-97,PolSavLeiSto-JCP-02,PolColLeiStoWerSav-IJQC-03,SavColPol-IJQC-03} to choose a modified interaction $v_{ee}^{\mu}(r)$ representing the long-range part of the Coulomb interaction. In previous works~\cite{Sav-INC-96,LeiStoWerSav-CPL-97,PolSavLeiSto-JCP-02,PolColLeiStoWerSav-IJQC-03}, this long-range interaction has been defined by
\begin{equation}
v_{ee,\erf}^{\mu}(r)=\frac{\erf(\mu r)}{r},
\label{veeerf}
\end{equation}
referred to as the erf interaction. More recently~\cite{TouColSav-PRA-04}, a sharper long-range/short-range separation has been achieved by taking for the long-range interaction
\begin{equation}
v_{ee,\erfgau}^{\mu}(r)=\frac{\erf(c \mu r)}{r} - \frac{2 c \mu}{\sqrt{\pi}} e^{-\frac{1}{3}c^2 \mu^2 r^2},
\label{veeerfgau}
\end{equation}
referred to as the erfgau interaction. This modified interaction was first introduced by Gill and Adamson~\cite{GilAda-CPL-96} in another context. In Eq.~(\ref{veeerfgau}), $c = \left( 1+6\sqrt{3}\right)^{1/2} \approx 3.375$ is a constant chosen so that $1/\mu$ roughly represents the range of the modified interaction as for the erf interaction~\cite{TouColSav-PRA-04}. For both the erf and erfgau interactions, the modified interaction $v_{ee}^{\mu}$ vanishes at $\mu=0$ and reduces to the full Coulomb interaction $1/r$ as $\mu \to \infty$. Consequently, the Hamiltonian $\hat{H}^{\mu}$ of Eq.~(\ref{Hmu}) is the KS non-interacting Hamiltonian at $\mu=0$ and the physical Hamiltonian as $\mu \to \infty$.

With these modified interactions, Eq.~(\ref{Hmu}) defines a fictitious system with long-range interactions from which one can define a universal long-range density functional
\begin{equation}
F^{\mu} = \bra{\Psimu} \hat{T} + \hat{V}_{ee}^{\mu} \ket{\Psimu},
\label{Umu}
\end{equation}
and its short-range complement $\bar{F}^{\mu} = F - F^{\mu}$ where $F$ is the usual Coulombic universal functional. This short-range universal functional $\bar{F}^{\mu}$ can be decomposed as
\begin{equation}
\bar{F}^{\mu} = \bar{U}^{\mu} + \bar{E}_{xc}^{\mu},
\label{Fsrmu}
\end{equation}
where $\bar{U}^{\mu}=1/2 \iint n(\b{r}_1) n(\b{r}_1) \bar{v}_{ee}^{\mu}(r_{12}) d\b{r}_1 d\b{r}_2$ is the complement short-range Hartree energy functional and $\bar{E}_{xc}^{\mu}$ is the unknown complement short-range exchange-correlation energy functional. Finally, the ground-state energy of the physical system is given by
\begin{equation}
E = \bra{\Psimu} \hat{T} + \hat{V}_{ee}^{\mu} \ket{\Psimu} + \int v_{ne}(\b{r}) n(\b{r}) d\b{r} + \bar{U}^{\mu} + \bar{E}_{xc}^{\mu},
\label{E}
\end{equation}
where $v_{ne}(\b{r})$ is the nuclei-electron potential. Accordingly, the potential $v^{\mu}$ appearing in Eq.~(\ref{Hmu}) can be decomposed as 
\begin{equation}
v^{\mu}(\b{r})=v_{ne}(\b{r})+v_{h}^{\mu}(\b{r})+v_{xc}^{\mu}(\b{r}),
\label{vmu}
\end{equation}
where $v_{h}^{\mu}(\b{r}) = \delta \bar{U}^{\mu}/\delta n(\b{r})$ is the complement short-range Hartree potential and $v_{xc}^{\mu}(\b{r})=\delta \bar{E}_{xc}^{\mu}/\delta n(\b{r})$ is the complement short-range exchange-correlation potential. As in the usual Kohn-Sham case, the short-range exchange-correlation energy can be decomposed into exchange and correlation contributions, $\bar{E}_{xc}^{\mu}=\bar{E}_{x}^{\mu}+\bar{E}_{c}^{\mu}$. Correspondingly, the short-range exchange-correlation potential is decomposed as $v_{xc}^{\mu}(\b{r})=v_{x}^{\mu}(\b{r}) + v_{c}^{\mu}(\b{r})$.

In this approach, the central quantity to approximate is the short-range exchange-correlation energy functional $\bar{E}_{xc}^{\mu}$. Once an approximation has been chosen for $\bar{E}_{xc}^{\mu}$, the potential $v^{\mu}$ appearing in Eq.~(\ref{Hmu}) is deduced by Eq.~(\ref{vmu}), the multi-determinantal wave function $\Psimu$ is calculated and the ground-state energy $E$ is deduced by Eq.~(\ref{E}). Previous applications of the method to small atomic and molecular systems~\cite{LeiStoWerSav-CPL-97,PolSavLeiSto-JCP-02,PedJen-JJJ-XX} show that, for a reasonable value of $\mu$, $\bra{\Psimu} \hat{T} + \hat{V}_{ee}^{\mu} \ket{\Psimu}$ can be efficiently approximated by standard wave function methods using a few-determinantal wave function $\Psimu$.

In practice, the short-range exchange-correlation energy is conveniently expressed as
\begin{equation}
\bar{E}_{xc}^{\mu} = \int n(\b{r})  \bar{\varepsilon}_{xc}^{\mu}(\b{r}) d\b{r}.
\label{}
\end{equation}
where $\bar{\varepsilon}_{xc}^{\mu}(\b{r})$ is a short-range local exchange-correlation energy per particle that is not uniquely defined. The local density approximation (LDA) for $\bar{\varepsilon}_{xc}^{\mu}$ is
\begin{equation}
\bar{\varepsilon}_{xc}^{\mu,\text{LDA}}(\b{r}) = \bar{\varepsilon}_{xc}^{\mu,\text{unif}}(n(\b{r})),
\label{epsxcLDA}
\end{equation}
where $\bar{\varepsilon}_{xc}^{\mu,\text{unif}}(n)$ is the short-range exchange-correlation energy per particle of a uniform electron gas with modified interaction~\cite{Sav-INC-96,TouSavFla-IJQC-04}. The LDA short-range exchange-correlation potential is
\begin{equation}
v_{xc}^{\mu,\text{LDA}}(\b{r}) = \left( \frac{d \left( n \, \bar{\varepsilon}_{xc}^{\mu,\text{unif}}(n) \right) }{d n} \right)_{n=n(\b{r})}.
\label{vxcLDA}
\end{equation}
The performance of LDA for $\bar{E}_{xc}^{\mu}$ has previously be investigated~\cite{LeiStoWerSav-CPL-97,PolSavLeiSto-JCP-02,PedJen-JJJ-XX,TouColSav-PRA-04}. It has been found in particular on atomic systems that the LDA is very accurate for sufficiently large $\mu$~\cite{TouColSav-PRA-04}. For example, for the He atom with $\mu \gtrsim 2$, the LDA error on both $\bar{E}_{x}^{\mu}$ and $\bar{E}_{c}^{\mu}$ is less than 1 mH. However, good accuracy on integrated quantities does not always ensure a similar accuracy on local quantities. In this work, we analyze further the LDA for short-range exchange-correlation effects by performing a detailed local analysis. Indeed, for both the erf and erfgau interactions, we have computed for the He and Be atoms accurate exchange-correlation potentials $v_{xc}^{\mu}$ and accurate local exchange-correlation energies per particle $\bar{\varepsilon}_{xc}^{\mu}$ in two possible definitions and compared to the corresponding LDA quantities.

The paper is organized as follows. In Sec.~\ref{sec:potential}, the calculation of accurate potentials $v^{\mu}$ is explained and the corresponding exchange-correlation potentials are compared to the LDA ones. In Sec.~\ref{sec:epsilons}, accurate local exchange-correlation energies per particle are calculated in two possible definitions and compared to the LDA local energy per particle. Finally, Sec.~\ref{sec:conclusion} contains concluding remarks.

Atomic units will be used throughout this work.

\section{Short-range exchange-correlation potentials}
\label{sec:potential}

The method we use for calculating accurate external potentials by optimization is explained in Refs.~\onlinecite{ColSav-JCP-99} and~\onlinecite{PolColLeiStoWerSav-IJQC-03}. We rapidly recall here the principal aspects.

The potentials $v^{\mu}(\b{r})$ are determined by using the Legendre transform formulation of the universal functional~\cite{Lie-IJQC-83,NalPar-JCP-82}
\begin{equation}
F^{\mu}[n] = \sup_{\tilde{v}^{\mu}} \Bigl\{ E^{\mu}[\tilde{v}^{\mu}] - \int \tilde{v}^{\mu}(\b{r}) n(\b{r}) d\b{r} \Bigl\},
\label{FmuLieb}
\end{equation}
where $E^{\mu}[\tilde{v}^{\mu}]$ is the ground-state energy of the Hamiltonian $\hat{T} + \hat{V}_{ee}^{\mu} + \sum_i \tilde{v}^{\mu}(\b{r}_i)$. If $n$ is chosen to be the physical density of the system, and if $n$ is assumed to be $v$-representable in the presence of the interaction $\hat{V}_{ee}^{\mu}$, then the supremum in Eq.~(\ref{FmuLieb}) is a maximum reached for the desired $v^{\mu}$ and $F^{\mu}$. In practice, an accurate density $n$ is computed by multi-reference configuration-interaction calculation with single and double excitations (MRCISD)~\cite{KnoWer-CPL-88,WerKno-JCP-88} and the potential to optimize $\tilde{v}^{\mu}(r)$ for atomic systems is expanded as
\begin{equation}
\tilde{v}^{\mu}(r) = \sum_{i=1}^{n} c_i r^{p_i} e^{\gamma_i r^2} + \frac{C}{r},
\label{vmuopt}
\end{equation}
where $c_i$ are the optimized coefficients, $p_i$ are some fixed integers ($-1$ or $2$),
$\gamma_i$ are fixed exponents chosen so as to form an even-tempered basis set (typically, $\gamma_i \in [10^{-3}, 5.10^4]$), and $C$ is a constant enforces the correct asymptotic behavior for $r \to \infty$. For the Kohn-Sham case ($\mu=0$), this asymptotic behavior is determined by the nuclei-electron, Hartree and exchange contribution to the potential, giving $C=-Z+N-1$ ($N$ and $Z$ are the electron number and nuclear charge, respectively). For finite $\mu$, the short-range Hartree and exchange potentials are exponentially decreasing at infinity, so that it remains only the nuclei-electron contribution, giving $C=-Z$. Actually, the maximizing potential $v^{\mu}(r)$ in Eq.~(\ref{FmuLieb}) is defined only up to an additive constant. The potential of Eq.~(\ref{vmuopt}) is the one which goes to $0$ as $r \to \infty$.

The maximization of Eq.~(\ref{FmuLieb}) is carried out with the Simplex method~\cite{PreTeuVetFla-BOOK-92}. For a given potential, $E^{\mu}[\tilde{v}^{\mu}]$ is computed at MRCISD
level using the Molpro program~\cite{Molpro-PROG-02} with modified two-electron integrals (see Appendix A of Ref.~\onlinecite{TouColSav-PRA-04}). Beside the asymptotic behavior for $r \to \infty$, $v^{\mu}(r) \sim C/r$, the behavior of the potential at the nucleus $r=0$, $v^{\mu}(r) \sim -Z/r$, is also imposed during the optimization. Large one-electron even-tempered Gaussian basis sets are used for all systems (see Refs.~\onlinecite{ColSav-JCP-99} and~\onlinecite{PolColLeiStoWerSav-IJQC-03} for more details). The quality of the obtained potential is assessed using the Zhao-Parr criterion~\cite{ZhaPar-PRA-92}
\begin{equation}
\label{}
\Delta = \frac{1}{2} \iint \frac{\left(\tilde{n}(\b{r}_1)-n(\b{r}_1) \right) \left(\tilde{n}(\b{r}_2)-n(\b{r}_2) \right)}{|\b{r}_1 - \b{r}_2|} d\b{r}_1 d\b{r}_2,
\end{equation}
where $\tilde{n}$ is the density given by the approximate potential $\tilde{v}^{\mu}$. The largest values of $\Delta$ obtained with our potentials are of order $10^{-8}$.

\begin{figure}
\includegraphics[scale=0.55]{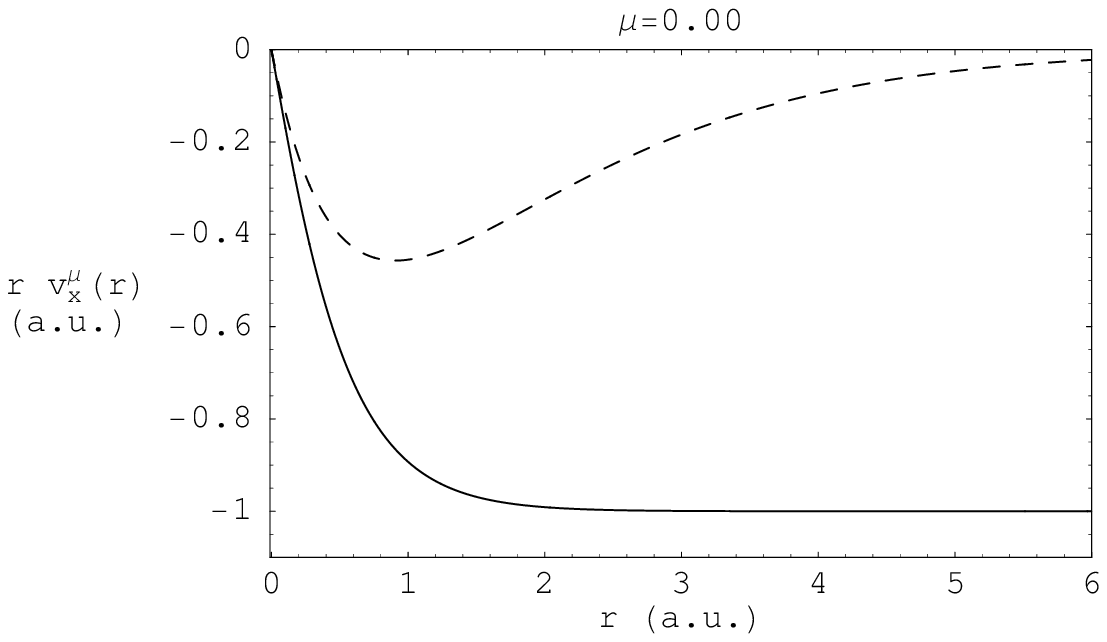}
\includegraphics[scale=0.55]{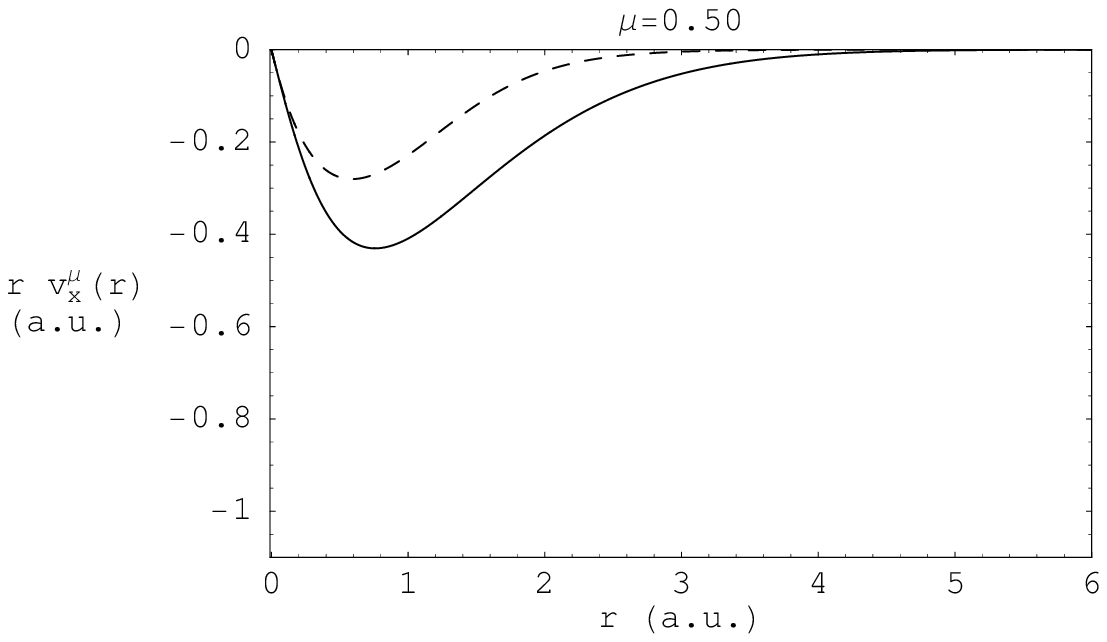}
\includegraphics[scale=0.55]{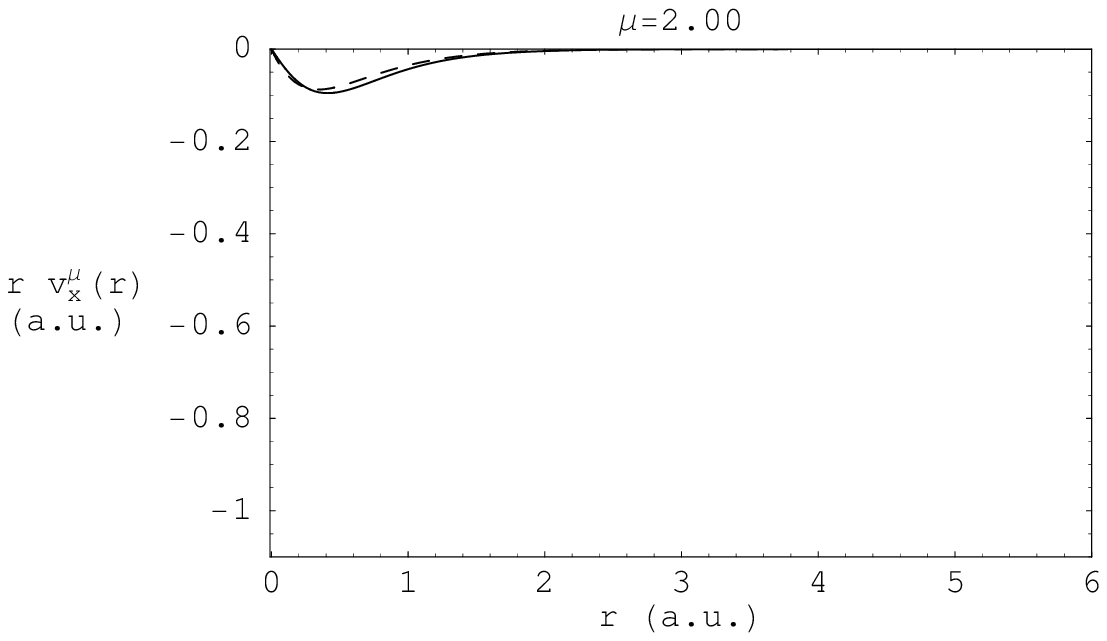}
\caption{Exchange potentials $r v_{x}^{\mu}(r)$ for He with the erf interaction with $\mu=0$, $0.5$ and $2$: the potentials obtained in the present work (full curves) are compared to the LDA potentials [Eq.~(\ref{vxcLDA}), dashed curves].
}
\label{fig:rvx-he-erf}
\end{figure}

For two-electron systems, the exchange-correlation potential can be easily decomposed into exchange and correlation contributions. Indeed, the exchange potential is known from the Hartree potential, $v_{x}^{\mu}(r) = - v_{h}^{\mu}(r)/2$, and the correlation potential is simply obtained by difference $v_{c}^{\mu}(r)=v_{xc}^{\mu}(r) - v_{x}^{\mu}(r)$. Figure~\ref{fig:rvx-he-erf} shows the exchange potentials of the He atom with the erf interaction for $\mu=0$, $0.5$ and $2$, together with the corresponding LDA potentials. To better visualize the asymptotic behavior as $r \to \infty$, $r v_{x}^{\mu}(r)$ is plotted instead of $v_{x}^{\mu}(r)$. In the KS case ($\mu=0$), the LDA exchange potential largely differs from the accurate one. In particular, the LDA potential has not the correct Coulombic asymptotic behavior $v_{x}^{\mu=0}(r) \sim -1/r$ as $r \to \infty$. When $\mu$ increases, the range of the accurate potentials decreases and $r v_{x}^{\mu}(r)$ goes to 0 as $r \to \infty$. When $\mu$ increases, the asymptote is reached at smaller and smaller values of $r$. The important observation is that the accuracy of the LDA potentials increases with $\mu$. At $\mu=2$, at the scale of the plot, the LDA potential is nearly identical to the accurate one.

\begin{figure}
\includegraphics[scale=0.55]{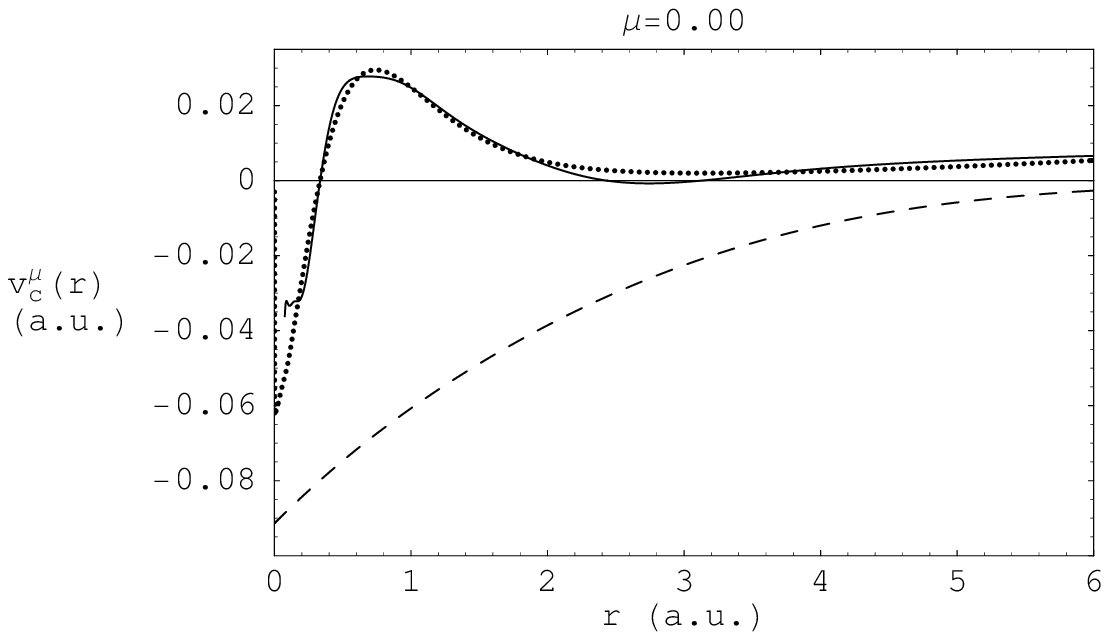}
\includegraphics[scale=0.55]{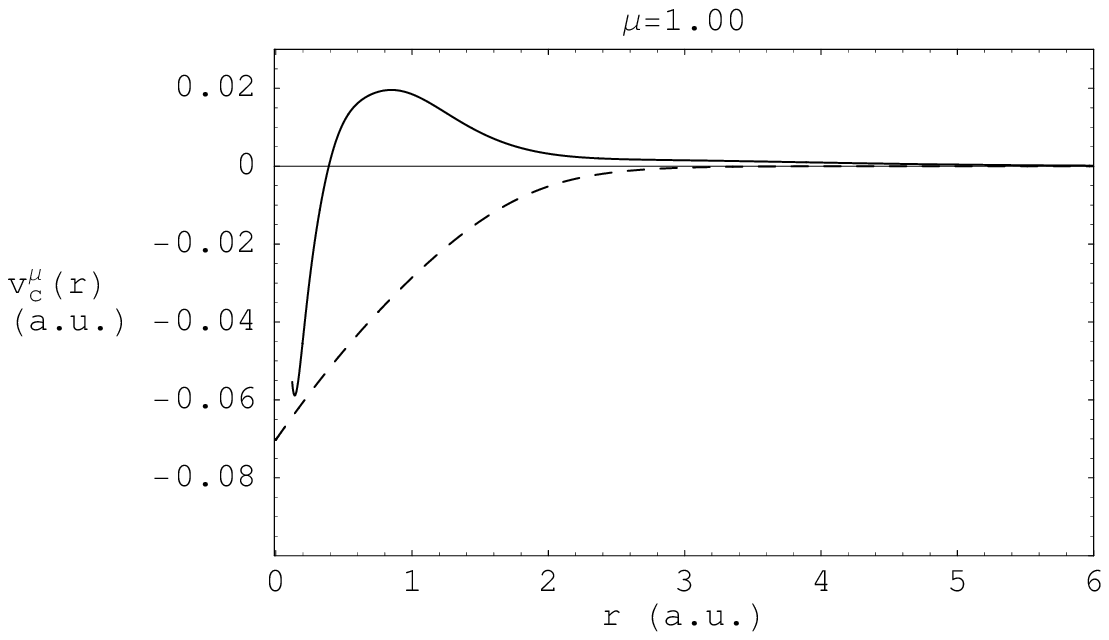}
\includegraphics[scale=0.55]{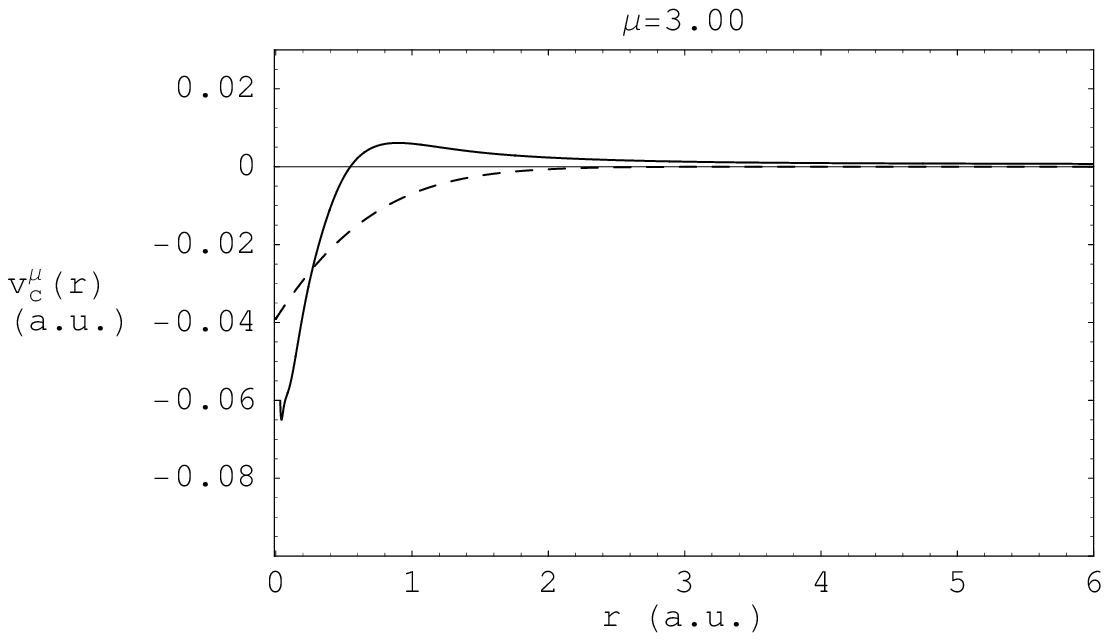}
\caption{Correlation potentials $v_{c}^{\mu}(r)$ for He with the erf interaction with $\mu=0$, $1$ and $3$: the potentials obtained in the present work (full curves) are compared to the LDA potentials [Eq.~(\ref{vxcLDA}), dashed curves]. For $\mu=0$, the correlation potential calculated by Umrigar and Gonze~\cite{UmrGon-PRA-94} (dotted curve) is also shown.
}
\label{fig:vc-he-erf}
\end{figure}

The accurate correlation potential for the He atom with the erf interaction for $\mu=0$, $1$ and $3$ is reported in figure~\ref{fig:vc-he-erf}, together with the LDA correlation potentials. In the KS case ($\mu=0$), the correlation potential calculated by Umrigar and Gonze~\cite{UmrGon-PRA-94} by inversion of the KS equation with a very accurate density is also reported. This potential agrees well with our potential except at very small $r$ where the precision of our calculations (using Gaussian basis sets) does not allow to extract the correlation potential. The LDA correlation potential for $\mu=0$ is a poor approximation to the accurate, structured potential. As for the exchange potential, the range of correlation potential is reduced when $\mu$ increases. At large $\mu$, the correlation potential has less structure which enables the LDA to perform better but on average only.

\begin{figure}
\includegraphics[scale=0.55]{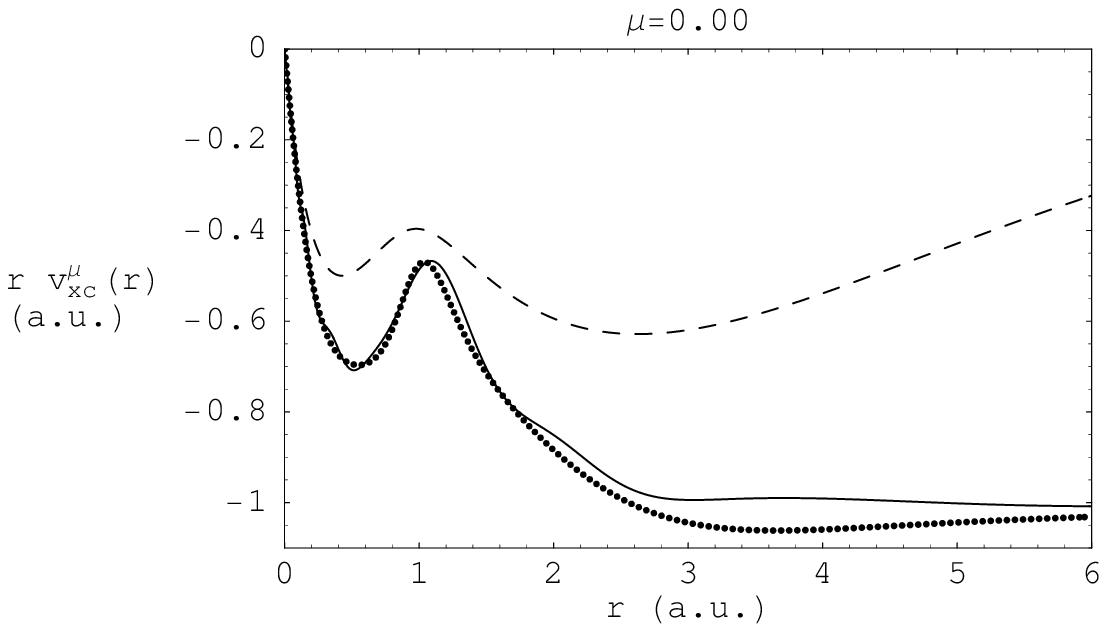}
\includegraphics[scale=0.55]{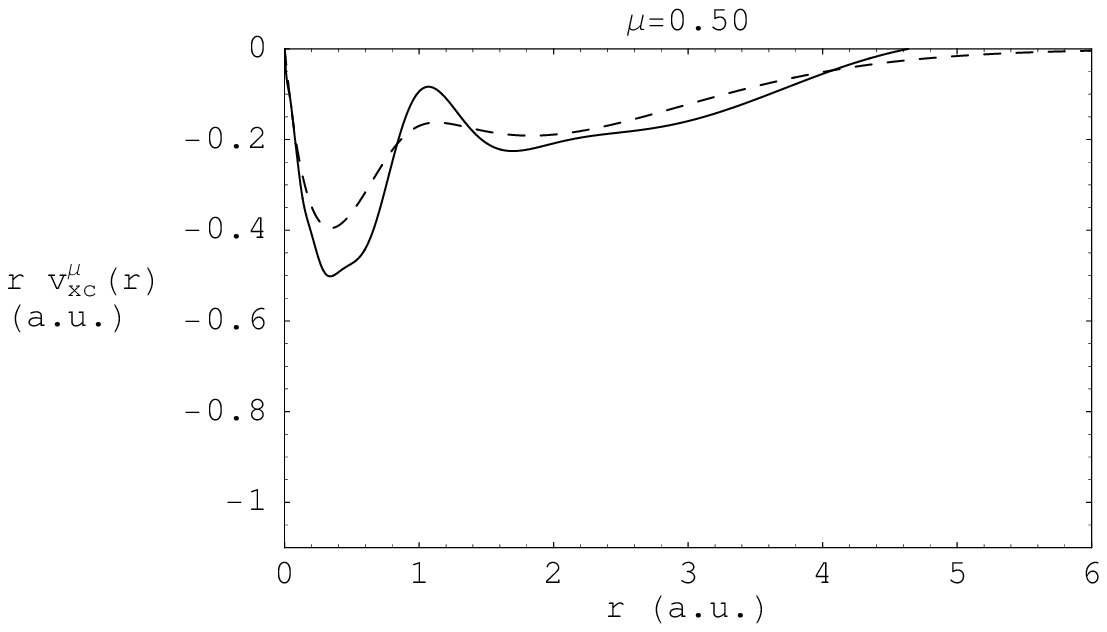}
\includegraphics[scale=0.55]{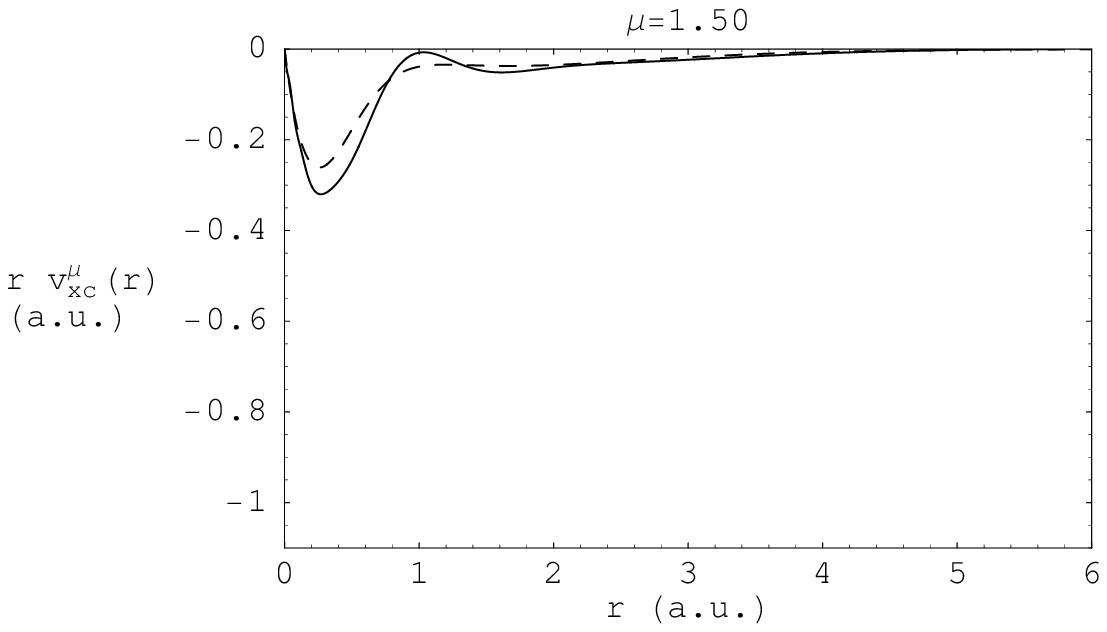}
\caption{Exchange-correlation potentials $r v_{xc}^{\mu}(r)$ for Be with the erf interaction with $\mu=0$, $0.5$ and $1.5$: the potentials obtained in the present work (full curves) are compared to the LDA potentials [Eq.~(\ref{vxcLDA}), dashed curves]. For $\mu=0$, the exchange-correlation potential calculated by Umrigar and Gonze~\cite{UmrGon-INC-93} (dotted curve) is also shown.
}
\label{fig:rvxc-be-erf}
\end{figure}

The potentials obtained for the Be atom with the erf interaction are plotted in figure~\ref{fig:rvxc-be-erf}. At $\mu=0$, the exchange-correlation potential calculated by Umrigar and Gonze~\cite{UmrGon-INC-93} using quantum Monte Carlo methods is also reported. In comparison to the He atom, the accurate potentials for small $\mu$ now exhibit a shell structure with a jump separating the core region ($r \lesssim 1$) and the valence region ($r \gtrsim 1$). The LDA potentials does not reproduce well this jump. When $\mu$ is large enough ($\mu \gtrsim 1$) so that the valence region is cut off, the shell structure disappears in the potential and the quality of the LDA potentials is improved.

The potentials with the erfgau interaction are qualitatively similar to the potentials with the erf interaction and will not be shown here. We just mention that a careful comparison shows that the potentials with the erfgau interaction are more short-ranged, confirming than the erfgau interaction achieves a better long-range/short-range separation than the erf interaction~\cite{TouColSav-PRA-04}.

\section{Short-range local exchange-correlation energies per particle}
\label{sec:epsilons}

The short-range local exchange-correlation energy per particle $\bar{\varepsilon}_{xc}^{\mu}(\b{r})$ associated to the exact $\bar{E}_{xc}^{\mu}$ is not uniquely defined. However, one can define and calculate some ``physically realistic'' $\bar{\varepsilon}_{xc}^{\mu}(\b{r})$ which can be compared to approximations like the LDA. We discuss now two such possible choices: local energies per particle obtained from integration of pair densities along adiabatic connections and local energies per particle obtained directly from potentials.

\subsection{Short-range local exchange-correlation energies per particle from pair densities}
\label{sec:epsxc}

It is natural to define a short-range local exchange energy per particle by
\begin{equation}
\bar{\varepsilon}_{x}^{\mu,\text{pd}}(\b{r}_1) = \frac{1}{2}  \int n_{x}(\b{r}_1,\b{r}_2) \bar{v}_{ee}^{\mu}(r_{12}) d\b{r}_2.
\label{epsxpd}
\end{equation}
where $n_x(\b{r}_1,\b{r}_2)$ is the exchange hole calculated from the one-particle KS density matrix: $n_x(\b{r}_1,\b{r}_2) = -|n_1^{\text{KS}}(\b{r}_1,\b{r}_2)|^2 / ( 2 n(\b{r}_1))$. The superscript pd stands for pair density. Similarly, the exact representation of $\bar{E}_{c}^{\mu}$ by integration over the non-linear adiabatic connection
\begin{equation}
\bar{E}_{c}^{\mu} = \frac{1}{2} \int_{\mu}^{\infty} d\xi \iint n(\b{r}_1) n_{c}^{\xi}(\b{r}_1,\b{r}_2) \frac{\partial v_{ee}^{\xi}(r_{12})}{\partial \xi} d\b{r}_1 d\b{r}_2,
\label{}
\end{equation}
where $n_{c}^{\xi}(\b{r}_1,\b{r}_2)$ is the correlation hole at interaction parameter $\xi$, suggests a natural definition for the short-range local correlation energy per particle (see also Ref.~\onlinecite{ColMaySav-PRA-03})
\begin{equation}
\bar{\varepsilon}_{c}^{\mu,\text{pd}}(\b{r}_1) = \frac{1}{2}  \int_{\mu}^{\infty} d\xi \int d\b{r}_2 n_{c}^{\xi}(\b{r}_1,\b{r}_2) \frac{\partial v_{ee}^{\xi}(r_{12})}{\partial \xi}.
\label{epscpd}
\end{equation}
Note that even for the KS system ($\mu=0$), the local correlation energy per particle $\bar{\varepsilon}_{c}^{\mu=0,\text{pd}}(\b{r})$ can depend on the adiabatic connection path followed. 

We have calculated these local energies per particle from accurate exchange and correlation holes associated to the system of Eq.~(\ref{Hmu}) using the accurate potentials of Sec.~\ref{sec:potential} and the programs Molpro~\cite{Molpro-PROG-02} and CASDI~\cite{BenMay-CPL-98}.

\subsection{Short-range local exchange-correlation energies per particle from potentials}
\label{sec:epsxclocal}

We now define another short-range local exchange-correlation energy per particle $\bar{\varepsilon}_{xc}^{\mu,\text{local}}(n)$ as a function of $n$, by requiring that it yields of course the exact short-range exchange-correlation energy
\begin{equation}
\bar{E}_{xc}^{\mu} = \int n(\b{r}) \bar{\varepsilon}_{xc}^{\mu,\text{local}}(n(\b{r})) d\b{r},
\label{Excfromlocal}
\end{equation}
but also the exact short-range exchange-correlation potential
\begin{equation}
\left( \frac{d \, \left( n \bar{\varepsilon}_{xc}^{\mu,\text{local}}(n) \right) }{d n} \right)_{n=n(\b{r})} = v_{xc}^{\mu}(\b{r}) + C.
\label{vxcdedn}
\end{equation}
The potential is defined only up to an additive constant $C$ and $v_{xc}^{\mu}(\b{r})$ is the exchange-correlation potential which goes to $0$ at infinity (calculated in Sec.~\ref{sec:potential}).

For a given system, $\bar{\varepsilon}_{xc}^{\mu,\text{local}}(n)$ defines an exact local short-range exchange-correlation functional in the sense that it gives the exact exchange-correlation potential and energy. This approach has already been applied for the Kohn-Sham scheme~\cite{SavCol-JMS-00a}; we briefly recall here how $\bar{\varepsilon}_{xc}^{\mu,\text{local}}(n)$ is calculated.

The condition of Eq.~(\ref{vxcdedn}) implies
\begin{equation}
\nabla \left( n \bar{\varepsilon}_{xc}^{\mu,\text{local}}(n) \right) = \frac{d \, \left( n \bar{\varepsilon}_{xc}^{\mu,\text{local}}(n) \right) }{d n} \nabla n = (v_{xc}^{\mu}(\b{r}) + C) \nabla n,
\label{}
\end{equation}
which becomes for spherically symmetric systems
\begin{equation}
\frac{d \left( n(r) \bar{\varepsilon}_{xc}^{\mu,\text{local}}(r) \right) }{d r} = \frac{d n(r) }{d r}  (v_{xc}^{\mu}(r) +C).
\label{dexcdr}
\end{equation}
Integration of Eq.~(\ref{dexcdr}) leads to
\begin{eqnarray}
n(r) \bar{\varepsilon}_{xc}^{\mu,\text{local}}(r) - \left( n(r) \bar{\varepsilon}_{xc}^{\mu,\text{local}}(r) \right)_{r \to \infty}=
\nonumber\\
 - \int_{r}^{\infty} \frac{d n(r')}{d r'} (v_{xc}^{\mu}(r') + C ) d r'.
\label{}
\end{eqnarray}
To avoid divergence of the integral in Eq.~(\ref{Excfromlocal}), we must have $\left( n(r) \bar{\varepsilon}_{xc}^{\mu,\text{local}}(r) \right)_{r \to \infty} = 0$; $\bar{\varepsilon}_{xc}^{\mu,\text{local}}(r)$ is then expressed as
\begin{equation}
\bar{\varepsilon}_{xc}^{\mu,\text{local}}(r) = \frac{-1}{n(r)} \int_{r}^{\infty} \frac{d n(r')}{d r'} v_{xc}^{\mu}(r') d r' + C ,
\label{epsxclocal}
\end{equation}
where the constant $C$ is fixed by requiring that
\begin{eqnarray}
\bar{E}_{xc}^{\mu} &=& \int_{0}^{\infty} n(r) \bar{\varepsilon}_{xc}^{\mu,\text{local}}(r) 4\pi r^2 d r
\nonumber\\
&=& -\frac{4\pi}{3} \int_{0}^{\infty} \frac{d \left( n(r) \bar{\varepsilon}_{xc}^{\mu,\text{local}}(r) \right)}{d r}r^3 d r
\nonumber\\
&=& -\frac{4\pi}{3} \int_{0}^{\infty} \frac{d n(r)}{d r} v_{xc}^{\mu}(r)  r^3 d r + C N,
\label{Exclocal}
\end{eqnarray}
with the number of electrons $N=\int n(r) 4\pi r^2 dr$. Similarly, using only the exchange or correlation parts of $v_{xc}^{\mu}(r)$ and $\bar{E}_{xc}^{\mu}$ in Eqs.~(\ref{epsxclocal}) and~(\ref{Exclocal}) leads to the local exchange energy per particle $\bar{\varepsilon}_{x}^{\mu,\text{local}}(r)$ and local correlation energy per particle $\bar{\varepsilon}_{c}^{\mu,\text{local}}(r)$, respectively. 

We have calculated these local energies per particle using the accurate potentials of Sec.~\ref{sec:potential} and accurate values of $\bar{E}_{x}^{\mu}$ and $\bar{E}_{c}^{\mu}$ obtained from $F^{\mu}$ of Eq.~(\ref{FmuLieb}).

\subsection{Results}

For monotonically decaying spherical densities, the map $r \to n(r)$ can be inverted and the local exchange-correlation energy per particle can therefore be expressed as a function of the density $n$ or, equivalently, as a function of $1/r_s=(4\pi n/3)^{1/3}$. The reason to use $1/r_s$ is that the exchange energy per particle of the Coulombic uniform electron gas $\bar{\varepsilon}^{\mu=0,\text{unif}}_{x}$ is proportional to $1/r_s$.

\begin{figure}
\includegraphics[scale=0.55]{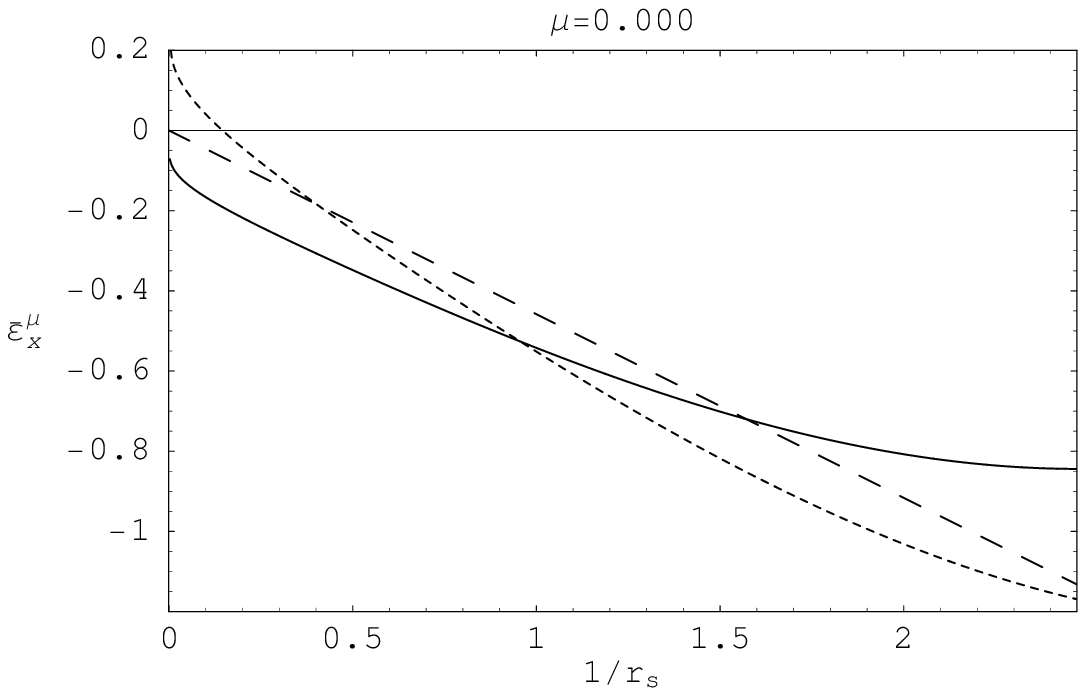}
\includegraphics[scale=0.55]{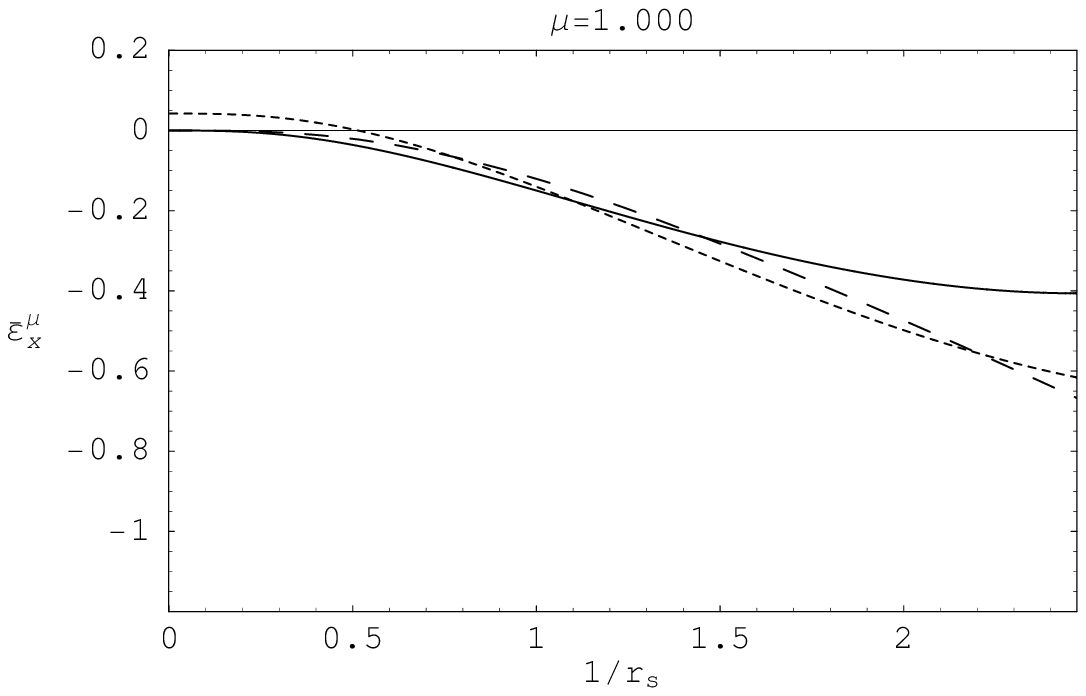}
\includegraphics[scale=0.55]{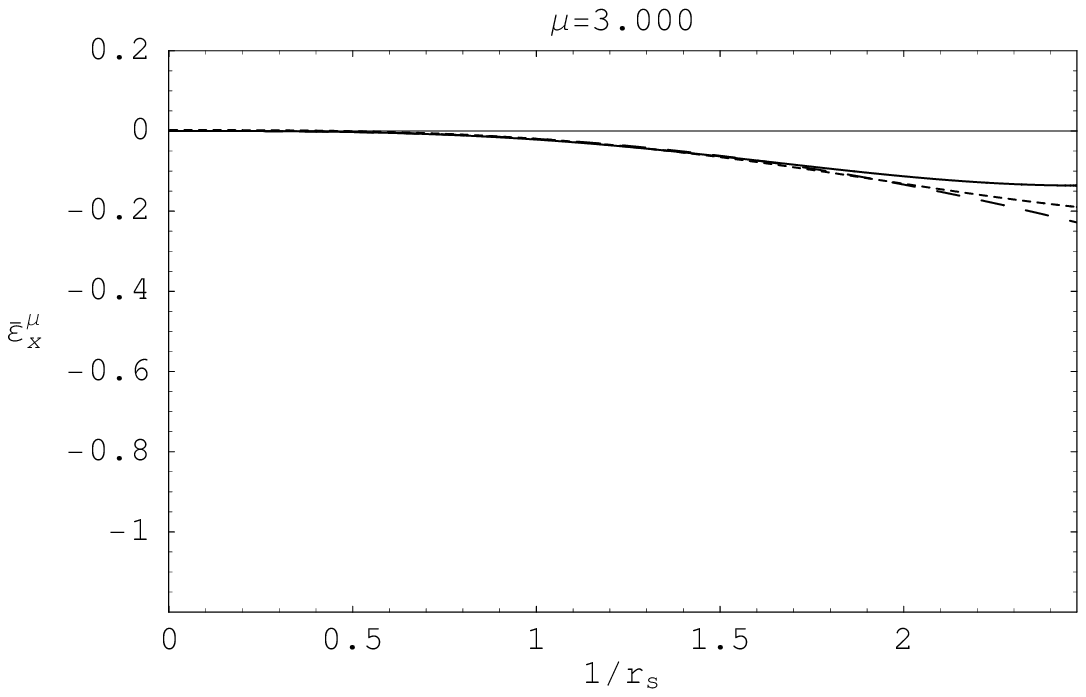}
\caption{Short-range local exchange energies per particle with respect to $1/r_s$ for He with the erf interaction for $\mu=0$, $1$, $3$: $\bar{\varepsilon}_{x}^{\mu,\text{LDA}}$ (long-dashed curve) is compared to the accurate $\bar{\varepsilon}_{x}^{\mu,\text{pd}}$ [Eq.~(\ref{epsxpd}), solid curve] and $\bar{\varepsilon}_{x}^{\mu,\text{local}}$ [Sec.~(\ref{sec:epsxclocal}), short-dashed curve].
}
\label{fig:epsx-he-erf-mu}
\end{figure}

For the He atom with the erf interaction, the two accurate short-range local exchange energies per particle $\bar{\varepsilon}_{x}^{\mu,\text{pd}}(1/r_s)$ and $\bar{\varepsilon}_{x}^{\mu,\text{local}}(1/r_s)$ are compared in figure~\ref{fig:epsx-he-erf-mu}, together with the LDA local energy per particle. The difference between the two accurate local energies per particle, both giving the same correct short-range exchange energy $\bar{E}_{x}^{\mu}$, illustrates the arbitrariness in the definition of local quantities from global ones. However, the differences soften when $\mu$ is increased, i.e. when long-range interactions are removed. The LDA local exchange energy per particle is significantly different from both accurate local exchange energies per particle at the KS end of the adiabatic connection ($\mu=0$), but as $\mu$ increases the LDA local energy per particle better and better agrees with the accurate local energies per particle. At low densities ($1/r_s < 0.5$), the LDA local energy per particle is close to the accurate $\bar{\varepsilon}_{x}^{\mu,\text{pd}}(1/r_s)$ while at high densities (for $1/r_s > 1.5$) the LDA local energy per particle is closer to the accurate $\bar{\varepsilon}_{x}^{\mu,\text{local}}(1/r_s)$.

\begin{figure}
\includegraphics[scale=0.55]{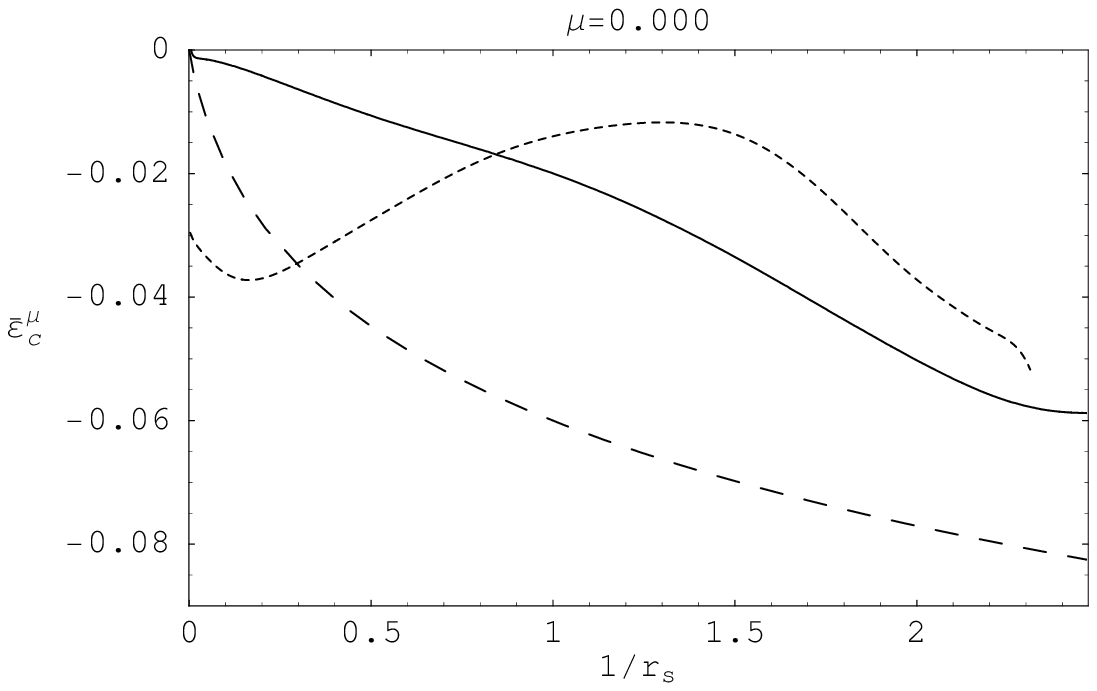}
\includegraphics[scale=0.55]{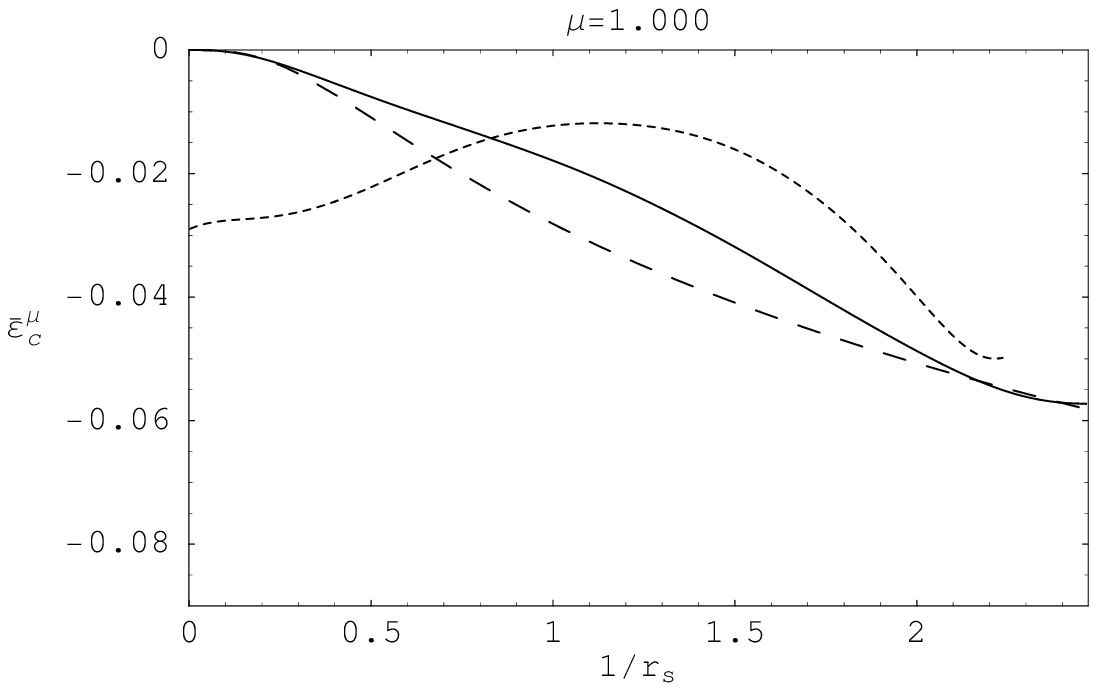}
\includegraphics[scale=0.55]{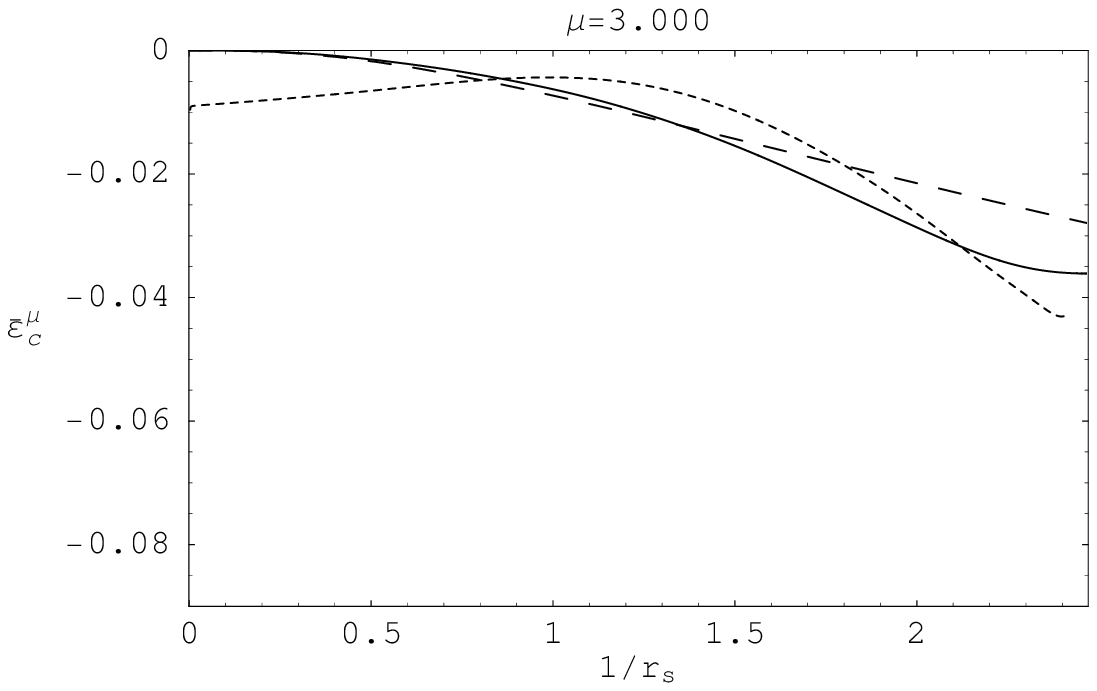}
\caption{Short-range local correlation energies per particle with respect to $1/r_s$ for He with the erf interaction for $\mu=0$, $1$, $3$. $\bar{\varepsilon}_{c}^{\mu,\text{LDA}}$ (long-dashed curve) is compared to the accurate $\bar{\varepsilon}_{c}^{\mu,\text{pd}}$ [Eq.~(\ref{epscpd}), solid curve] and $\bar{\varepsilon}_{c}^{\mu,\text{local}}$ [Sec.~(\ref{sec:epsxclocal}), short-dashed curve].
}
\label{fig:epsc-he-erf-mu}
\end{figure}

In figure~\ref{fig:epsc-he-erf-mu}, again for the He atom with the erf interaction, the accurate local correlation energies per particle $\bar{\varepsilon}_{c}^{\mu,\text{pd}}(1/r_s)$ and $\bar{\varepsilon}_{c}^{\mu,\text{local}}(1/r_s)$ are represented and compared to the LDA. The two accurate local energies per particle give the same correct short-range correlation energy $\bar{E}_{c}^{\mu}$ but have rather different shapes, although the differences partly soften when $\mu$ increases. The LDA generally overestimates the accurate local correlation energies per particle at $\mu=0$, but again when $\mu$ increases the LDA is improved on average. The LDA local energy per particle is much more comparable to $\bar{\varepsilon}_{c}^{\mu,\text{pd}}(1/r_s)$ than $\bar{\varepsilon}_{c}^{\mu,\text{local}}(1/r_s)$ especially for large $\mu$ and at low densities. 


\begin{figure}
\includegraphics[scale=0.55]{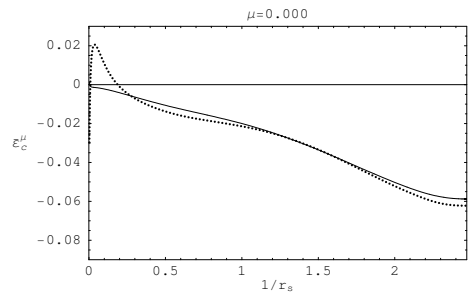}
\caption{Accurate local short-range correlation energies per particle $\bar{\varepsilon}_{c}^{\mu,\text{pd}}$ [Eq.~(\ref{epscpd})] with respect to $1/r_s$ for He with the erf (solid curve) and erfgau (dotted curve) interactions for $\mu=0$.
}
\label{fig:epsc-he-erferfgau-mu0.000}
\end{figure}

In general, similar behaviors are obtained with the erfgau interaction. However, the attractive character of this interaction for small values of $\mu$ (see Ref.~\onlinecite{TouSavFla-IJQC-04}) can lead to some significant differences compared to the erf interaction. Figure~\ref{fig:epsc-he-erferfgau-mu0.000} shows these differences in the accurate local correlation energy per particle $\bar{\varepsilon}_{c}^{\mu,\text{pd}}(1/r_s)$ at $\mu=0$. In comparison to the erf interaction, the local energy per particle obtained with the erfgau interaction exhibits in particular a positive contribution at very low densities. This emphasizes now the arbitrariness of the definition $\bar{\varepsilon}_{c}^{\mu,\text{pd}}(1/r_s)$ with respect to the choice of the adiabatic connection along which the pair density is integrated.

\begin{figure}
\includegraphics[scale=0.55]{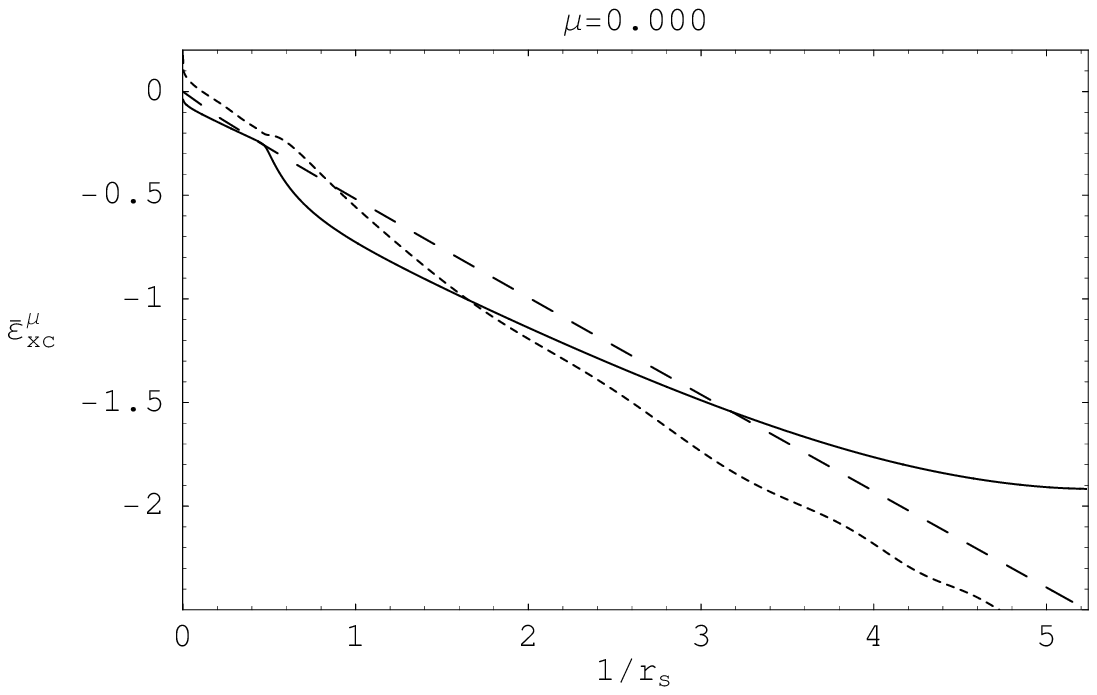}
\includegraphics[scale=0.55]{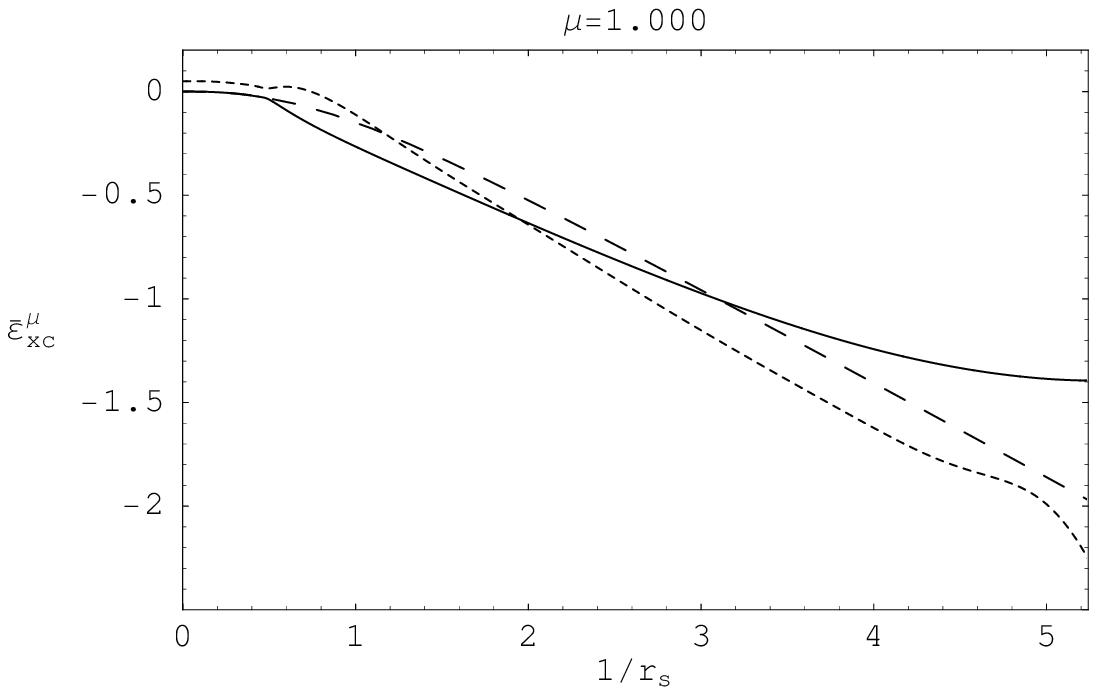}
\includegraphics[scale=0.55]{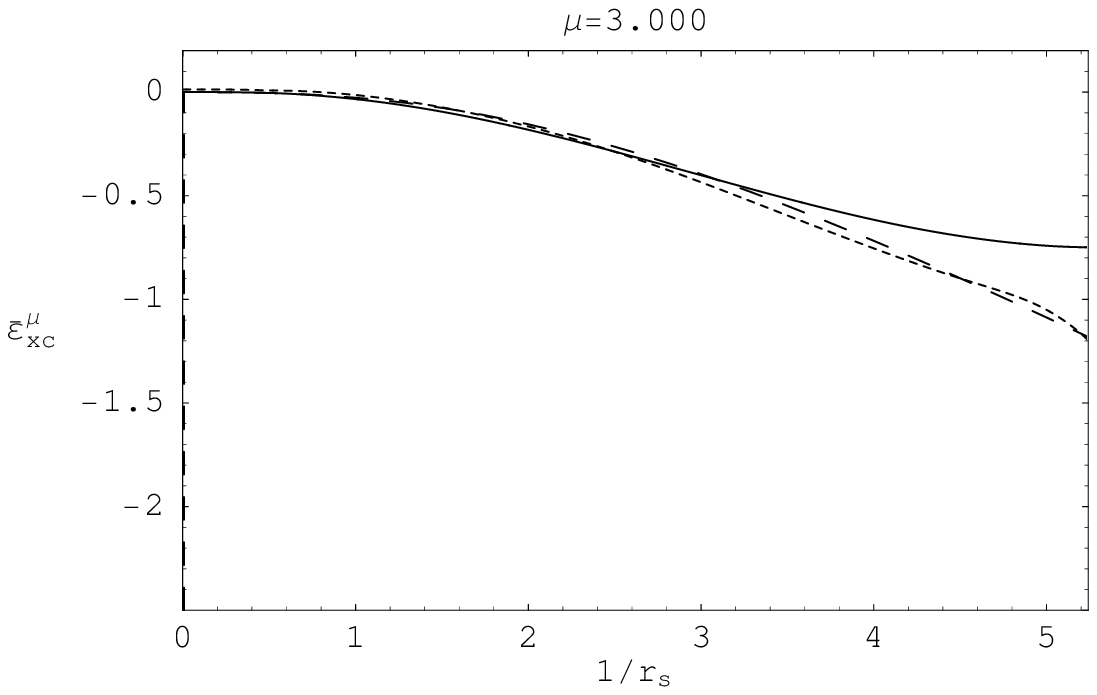}
\caption{Short-range local exchange-correlation energies per particle with respect to $1/r_s$ for Be with the erf interaction for $\mu=0$, $1$, $3$: $\bar{\varepsilon}_{xc}^{\mu,\text{LDA}}$ [Eq.~(\ref{epsxcLDA}), long-dashed curve] is compared to the accurate $\bar{\varepsilon}_{xc}^{\mu,\text{pd}}$ [Sec.~(\ref{sec:epsxc}), solid curve] and $\bar{\varepsilon}_{xc}^{\mu,\text{local}}$ [Eq.~(\ref{epsxclocal}), short-dashed curve].
}
\label{fig:epsxc-be-erf-mu}
\end{figure}

Figure~\ref{fig:epsxc-be-erf-mu} show the evolution of the local exchange-correlation energies per particle with the erf interaction for the Be atom. In comparison to the He atom, both $\bar{\varepsilon}_{c}^{\mu,\text{pd}}(1/r_s)$ and $\bar{\varepsilon}_{c}^{\mu,\text{local}}(1/r_s)$ exhibit a shell structure for small values of $\mu$, the contributions from the valence region ($r_s \lesssim 0.5$) and the core region ($r_s \gtrsim 0.5$) being easily identifiable. The LDA local exchange-correlation energy per particle differs significantly from the two accurate ones, especially in the core region which constitutes the major contribution to the exchange-correlation energy. Also, in this representation in term of $1/r_s$, the LDA does not reproduce at all the intershell jump. However, when $\mu$ increases, the contribution from the valence region is progressively cut off and the shell pattern finally disappears. For large $\mu$, when only the contribution from the core remains, the LDA local energy is close to the accurate local energies.

\begin{figure}
\includegraphics[scale=0.55]{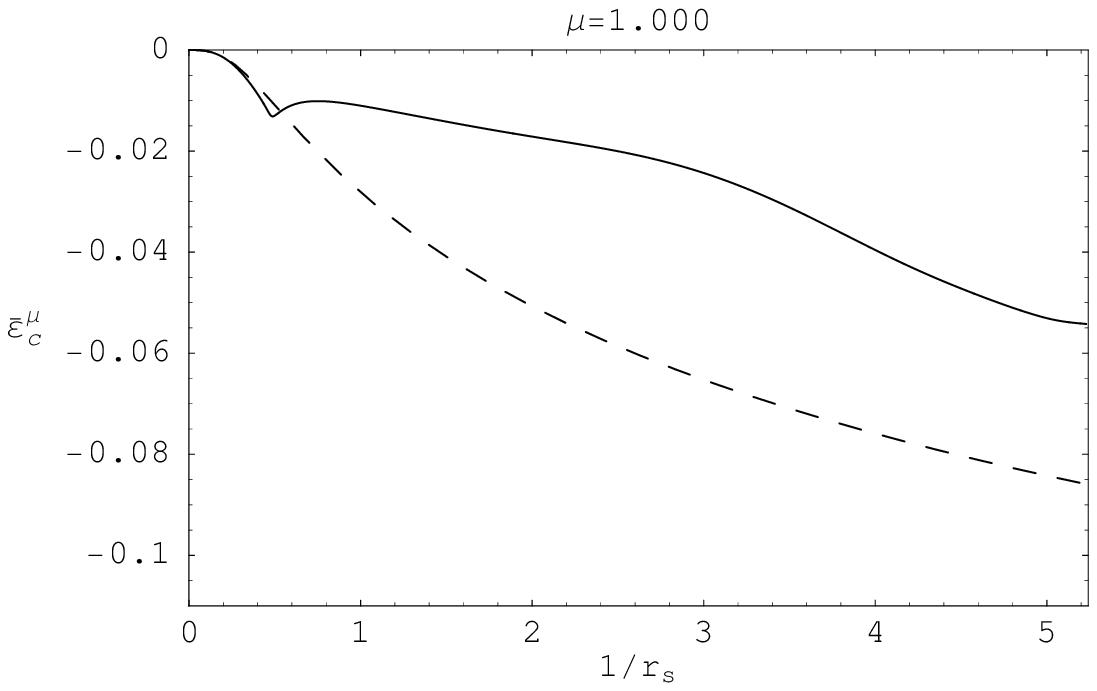}
\includegraphics[scale=0.55]{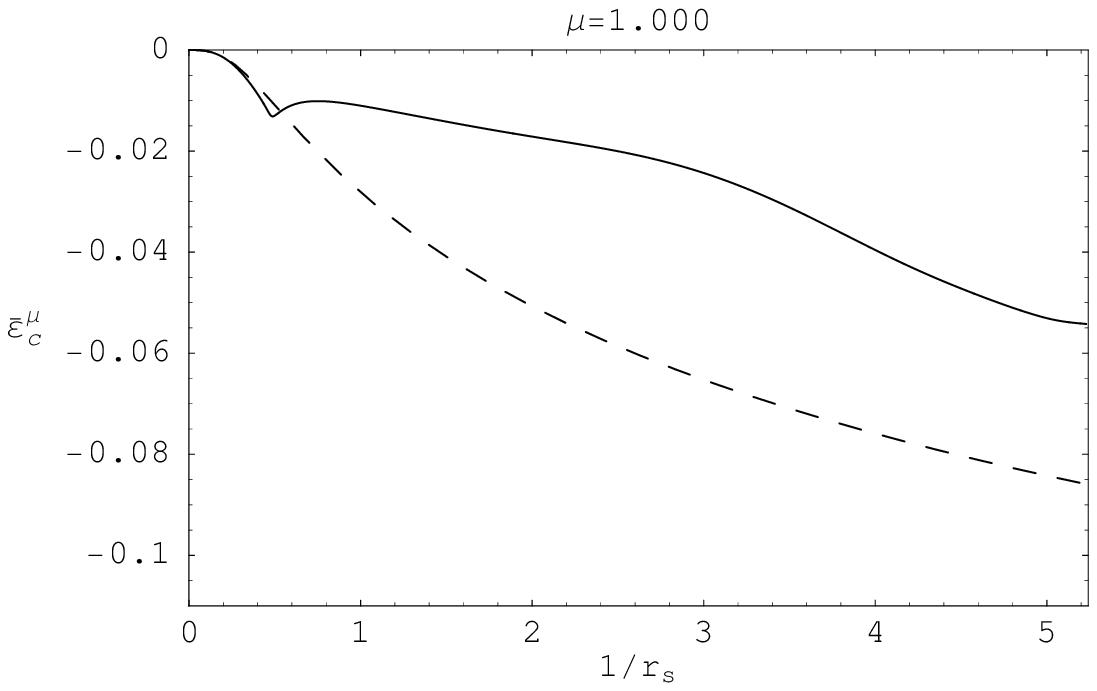}
\includegraphics[scale=0.55]{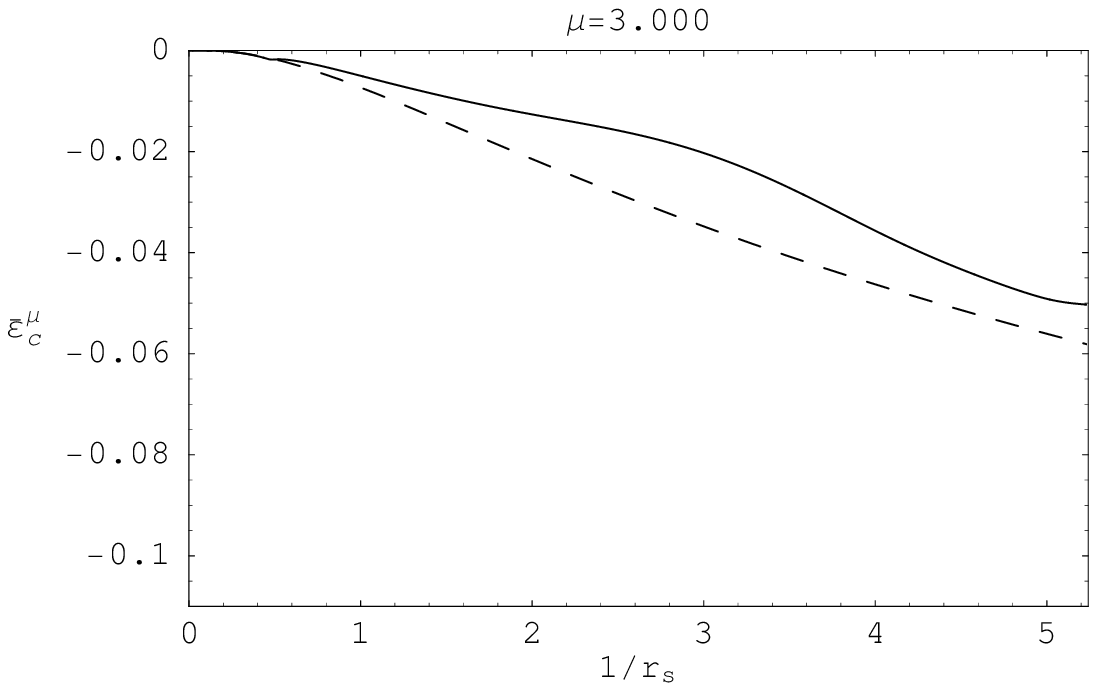}
\caption{Short-range local correlation energies per particle with respect to $1/r_s$ for Be with the erf interaction for $\mu=0$, $1$, $3$: $\bar{\varepsilon}_{c}^{\mu,\text{LDA}}$ (long-dashed curve) is compared to the accurate $\bar{\varepsilon}_{c}^{\mu,\text{pd}}$ [Eq.~(\ref{epscpd}), solid curve].
}
\label{fig:epsc-be-erf-mu}
\end{figure}

Finally, the accurate local correlation energy per particle $\bar{\varepsilon}_{c}^{\mu,\text{pd}}(1/r_s)$ for the Be atom is reported in figure~\ref{fig:epsc-be-erf-mu}, together with the LDA. The shell structure in this plot is even clearer. At $\mu=0$, the LDA overestimates the accurate local correlation energy per particle at all densities. When $\mu$ increases, the LDA starts to better reproduce the valence region but still overestimate the accurate local energy in the core region. Keeping on increasing $\mu$, the contribution from the valence region vanishes, and the LDA starts to reproduce well the core region.

\section{Concluding remarks}
\label{sec:conclusion}

Non-linear adiabatic connections have been used to decompose the energy of an electronic system into a long-range wave function part and short-range density functional part, the position of the frontier of this decomposition being determined by a single parameter $\mu$.

When the interaction range of the short-range energy component is reduced, corresponding to a rise of $\mu$, the short-range LDA becomes an increasingly accurate approximation to the short-range exchange-correlation density functional. Indeed, as it has been verified in this work for a few atomic systems, the short-range LDA gives, for a sufficiently large $\mu$, good short-range exchange-correlation potentials and local exchange-correlation energies per particle.

More formally, one can define an optimal local interaction parameter, $\mu_{\text{opt}}(\b{r})$, so that for $\mu > \mu_{\text{opt}}(\b{r})$ the LDA reproduces, at a given precision, the exact short-range potentials or local energies at the considered point of space. Physically, $1/\mu_{\text{opt}}(\b{r})$ represents the maximum local interaction range over which the exchange-correlation effects are well transferable from the uniform electron gas to the inhomogeneous system of interest. In general, the larger the density at point $\b{r}$ is, the larger $\mu_{\text{opt}}(\b{r})$ is. However, it is not conceivable in the method to use a $\b{r}$-dependent interaction parameter $\mu$ because of the wave function part of the calculation.

In a simple system like the He atom, the improvement of the LDA upon increasing $\mu$ is quite uniform in space and one can basically define an global average optimal interaction parameter, $\bar{\mu}_{\text{opt}}$, and use it to fix the frontier of the decomposition, i.e. $\mu = \bar{\mu}_{\text{opt}}$.

More generally, in strongest inhomogeneous systems like the Be atom, at least two regions of space are identifiable, the core and valence shells, and it is meaningful to define separate averages of $\mu_{\text{opt}}(\b{r})$ over each regions, $\bar{\mu}_{\text{opt}}^{\text{core}}$ and $\bar{\mu}_{\text{opt}}^{\text{valence}}$. As the density is higher in the core than in the valence, we have $\bar{\mu}_{\text{opt}}^{\text{core}} > \bar{\mu}_{\text{opt}}^{\text{valence}}$. 

If one fixes the frontier of the long-range/short-range decomposition at the optimal value for the core shell, i.e. $\mu = \bar{\mu}_{\text{opt}}^{\text{core}}$, the short-range functional part of the calculation can be well treated in LDA but of course the complement part of the core region and the entire valence region are assigned to the wave function part which can result in the need of a long expansion into Slater determinants.

On the contrary, if one fixes the frontier of the decomposition at the optimal value for the valence shell, i.e. $\mu = \bar{\mu}_{\text{opt}}^{\text{valence}}$, the short-range LDA functional is not enable to treat well the core region but can describe well that part of the valence region assigned to it. Moreover, the wave function part of the calculation has only to handle with the remaining part of the valence region which allows in general the use a short expansion into Slater determinants.

Obviously, one would like to re-conciliate these two alternatives. For practical applications of the method to molecular systems of chemical interests, it is desirable to keep the wave function expansion minimal and thus to chose the frontier of the decomposition inside the valence region, i.e. $\mu = \bar{\mu}_{\text{opt}}^{\text{valence}}$. For properties essentially depending on the valence shell only, such as atomization energies, the error of the LDA in the core shell can then be ignored, or the core electrons can be replaced by a pseudo-potential. Otherwise, one must then find better approximations to the short-range exchange-correlation functional which extend the domain of accuracy of the LDA toward larger interaction ranges (small $\mu$). We hope that the presented results will help the construction of such approximations.

Data for the atomic systems presented in this paper (and other systems) are available from the authors upon request.

\begin{acknowledgments}
We thank D. Maynau (Universit\'e Paul Sabatier, Toulouse, France) for providing us the CASDI program. We are also grateful to C. Umrigar (Cornell University, USA) for the data of Refs.~\onlinecite{UmrGon-PRA-94,UmrGon-INC-93}.
\end{acknowledgments}

\bibliographystyle{apsrev}
\bibliography{biblio}

\end{document}